\documentclass[%
 aps,reprint,
 superscriptaddress,
 amsmath,amssymb
]{revtex4-2}
\usepackage{placeins}
\usepackage{graphicx}
\usepackage{dcolumn}
\usepackage{bm}
\usepackage{xspace}
\usepackage{booktabs}  

\usepackage{xcolor}
\usepackage{longtable}
\begin{document}


\title{Exploring the $t\bar{t}$ threshold at an electron-positron collider}

\author{Leyan Li}
\author{Yuming Lin}
\author{Xiaohu Sun}
\email[Corresponding author, ]{Xiaohu.Sun@pku.edu.cn}
\author{Yajun Mao}
\affiliation{School of Physics and State Key Laboratory of Nuclear Physics and Technology, Peking University, Beijing 100871, China}%
\affiliation{Center for High Energy Physics, Peking University, Beijing 100871, China}

\author{Zhan Li}
\author{Kaili Zhang}
\thanks{Deceased.}
\author{Shudong Wang}
\author{Gang Li}
\author{Hongbo Liao}
\author{Yaquan Fang}
\thanks{Deceased.}
\affiliation{Institute of High Energy Physics, Chinese Academy of Sciences, Beijing 100049, China}%
\affiliation{University of Chinese Academy of Sciences, Beijing 100049, China}



\begin{abstract}
Future electron-positron colliders offer a unique opportunity for high-precision measurements of the top-quark mass, width, strong coupling constant, and top-quark Yukawa coupling via a scan of the $t\bar{t}$ threshold. We present the first prospect study of the simultaneous determination of these parameters, incorporating the latest reference detector design for the Circular Electron-Positron Collider (CEPC). We find that the precision of the top-quark mass measurement can reach a few MeV excluding the theoretical uncertainty on the cross-section, which is nearly two orders of magnitude better than the high-luminosity LHC (HL-LHC) projections. The current theoretical uncertainty of the cross-section calculation is the limiting factor.
\end{abstract}

\maketitle

\newcommand{\mt}{\ensuremath{m_{\text{t}}}\xspace}
\newcommand{\mtmc}{\ensuremath{m_{\text{t}}^{\text{MC}}}\xspace}
\newcommand{\mtpole}{\ensuremath{m_{\text{t}}^{\text{Pole}}}\xspace}
\newcommand{\mtmsbar}{\ensuremath{m_{\text{t}}^{\overline{\text{MS}}}}\xspace}

\newcommand{\wt}{\ensuremath{\Gamma_{\text{t}}}\xspace}
\newcommand{\ifb}{\ensuremath{\text{fb}^{-1}}\xspace}
\newcommand{\lumiN}{\ensuremath{56~\text{fb}^{-1}}\xspace}
\newcommand{\lumiU}{\ensuremath{84~\text{fb}^{-1}}\xspace}
\newcommand{\alphas}{\ensuremath{\alpha_{\text{S}}}\xspace}
\newcommand{\yt}{\ensuremath{y_{\text{t}}}\xspace}
\newcommand{\ttbar}{\ensuremath{t\bar{t}}\xspace}
\newcommand{\sqrts}{\ensuremath{\sqrt{s}}\xspace}

\newcommand{\WW}{\ensuremath{WW}\xspace}
\newcommand{\ZWW}{\ensuremath{ZWW}\xspace}
\newcommand{\ZZ}{\ensuremath{ZZ}\xspace}
\newcommand{\ZZZ}{\ensuremath{ZZZ}\xspace}
\newcommand{\bbbar}{\ensuremath{b\bar{b}}\xspace}
\newcommand{\qqbar}{\ensuremath{q\bar{q}}\xspace}

\newcommand{\pt}{\ensuremath{p_{\text{T}}}\xspace}
\newcommand{\met}{\ensuremath{E_\text{T}^{\text{miss}}}\xspace}

The top quark, discovered at the Fermilab Tevatron in 1995~\cite{PhysRevLett.74.2626,PhysRevLett.74.2632}, is the heaviest elementary particle in the Standard Model (SM). Its mass, \mt, originating mostly from the strongest Yukawa coupling to the Higgs boson, is fundamental to the consistency of the SM~\cite{ALEPH:2010aa,deBlas:2022hdk} and governs the stability of the electroweak vacuum through quantum corrections to the Higgs potential~\cite{DESIMONE20091,BEZRUKOV2008703,Bezrukov:2012sa,Degrassi:2012ry}.

To date, \mt has been measured only at hadron colliders via direct reconstruction of top-quark decay products. The combined Tevatron result~\cite{CDF:2012hvh,CDF:2016vzt} is \mt = 174.30 $\pm$ 0.35 (stat.) $\pm$ 0.54 (syst.)~GeV with a total uncertainty of 650~MeV, while the LHC combination~\cite{ATLAS:2024dxp} yields \mt = 172.52 $\pm$ 0.14 (stat.) $\pm$ 0.30 (syst.)~GeV with a total uncertainty of 330~MeV. The HL-LHC projection anticipates a precision of 200~MeV~\cite{ECFA16,ATLAS:2025jbw}. These measurements correspond to \mtmc, a generator-level parameter~\cite{Nason:2017cxd,Hoang:2020iah}, and are ultimately limited by non-perturbative effects at the scale $\Lambda_{\text{QCD}} \sim 200$~MeV. The precision from hadron colliders is insufficient to determine whether the electroweak vacuum is stable or meta-stable.

The generator mass \mtmc is neither a well-defined field-theoretic parameter nor rigorously related to the mass defined under a certain renormalization scheme, such as \mtpole and \mtmsbar that are convertible at four-loop accuracy~\cite{Marquard:2015qpa,Nason:2017cxd,Hoang:2020iah}. The extraction of \mtpole using the cross-section at the LHC and its projection to HL-LHC reach only GeV-level precision~\cite{CMS:2016yys,ECFA16}.

At electron-positron colliders, the \ttbar cross-section near threshold is highly sensitive to \mt, enabling a threshold-scan measurement with $\mathcal{O}(10)$~MeV precision in a well-defined renormalization scheme, free of $\Lambda_{\text{QCD}}$ ambiguities~\cite{Fadin:1987wz,Fadin:1988fn,Strassler:1990nw,Bigi:1986jk,Nason:2017cxd,Hoang:2020iah}. Previous studies cover ILC, CLIC, CEPC and FCC-ee~\cite{Martinez:2002st,Seidel:2013sqa,Horiguchi:2013wra,CLICdp:2018esa,FCC:2018evy,Fcc-ee2025,Li:2022iav}.

This work for the first time takes the latest CEPC reference detector~\cite{CEPCStudyGroup:2025kmw} to evaluate the threshold-scan prospects, simultaneously extracting \mt, \wt, \alphas and the top Yukawa coupling modifier \yt (relative scale, e.g. \yt=1.0 for the SM). Dedicated kinematic selections are newly optimized with simulation at the \ttbar threshold for CEPC and uncertainties are evaluated via a profiled likelihood scan, except the theoretical uncertainty of \ttbar cross-section from scale variations which is not profiled.

The \ttbar cross-section at the threshold is calculated at N$^3$LO QCD with resummed non-relativistic perturbation theory~\cite{Beneke:2013jia,Beneke:2015kwa} and NNLO in electroweak~\cite{Beneke:2017rdn}, using QQbar\_threshold (v2.2.0)~\cite{Beneke:2016kkb,Beneke:2017rdn}. The potential-subtracted (PS) scheme~\cite{Beneke:1998rk} is adopted to define the top-quark mass, which avoids the $\mathcal{O}(\Lambda_{\text{QCD}})$ renormalon ambiguity inherent in the pole mass and yields a more stable perturbative expansion near threshold. The nominal values are \mt = 171.5~GeV, \wt = 1.33~GeV, \alphas = 0.1184 and \yt = 1.0.

The \ttbar cross-section as a function of \sqrts near threshold is shown in Fig.~\ref{fig:xs}. The N$^3$LO curve is the baseline calculation from QQbar\_threshold. The N$^3$LO+ISR curve includes the initial-state radiation (ISR) effect at leading logarithmic (LL) precision; ISR photons carry away beam energy, reducing the cross-section. The N$^3$LO+ISR+LS curve furthermore accounts for the luminosity spectrum (LS) arising from beam energy spread, modeled by convoluting a Gaussian energy distribution with width $\sigma_{\text{LS}} = 0.51~\text{GeV} \times (\sqrts / 360~\text{GeV})^2$. The LS smears the cross-section enhancement at threshold, but this effect is significantly less severe in circular colliders such as CEPC and FCC-ee, where bending magnets constrain the energy spread, compared to linear colliders~\cite{Li:2022iav}.



The sensitivity to the physical parameters $\theta \in \{\mt,\wt,\alphas,\yt\}$ is quantified via the Fisher information. For an expected event yield $n_0 = \mathcal{L} \sigma_0$, where $\mathcal{L}$ denotes the integrated luminosity and $\sigma_0$ is the theoretical cross-section, the observed count $n$ is assumed to follow a Gaussian distribution $G(n \mid n_0, \sqrt{n_0})$ as a large-statistics approximation of Poisson fluctuations. The Fisher information $I(\sqrt{s})$ is obtained by taking the expectation of the squared logarithmic derivative of the likelihood:

\begin{align}
\begin{split}
I(\sqrt{s}) &= \int \left( \frac{\partial \log G(n \mid n_{0}, \sqrt{n_{0}})}{\partial\theta} \right)^{2} \\
& \quad \times G(n \mid n_{0}, \sqrt{n_{0}}) \, dn
\end{split}
\end{align}



Fig.~\ref{fig:fisher} presents the normalized Fisher information for $\theta \in \{\mt,\wt,\alphas,\yt\}$, derived from the N$^3$LO+ISR+LS cross-section in Fig.~\ref{fig:xs}. Peak positions dictate the optimal energy for each parameter: the threshold region maximizes sensitivity to \mt, while the cross-section peak is most sensitive to \wt (which broadens the resonance) as well as \alphas and \yt (which govern the overall rate). To simultaneously capture all information peaks, a five-point scan at \sqrts = \{342.0, 342.5, 343.0, 343.5, 344.0\}~GeV is chosen.

The instantaneous luminosity at the \ttbar threshold is expected to reach $1.1 \times 10^{34}$~cm$^{-2}$s$^{-1}$ per interaction point with a synchrotron radiation power of 50~MW~\cite{CEPCStudyGroup:2025kmw}. One year of data-taking from two interaction points yields 280~\ifb, defined as the baseline scenario (BASE). A 50\% luminosity upgrade to 420~\ifb defines the extended scenario (EXT), consistent with FCC-ee estimates~\cite{Fcc-ee2025}. The luminosity per scan point is thus 56 (84)~\ifb for BASE (EXT).

\begin{figure}[htbp]
  \centering
  \includegraphics[scale=0.30]{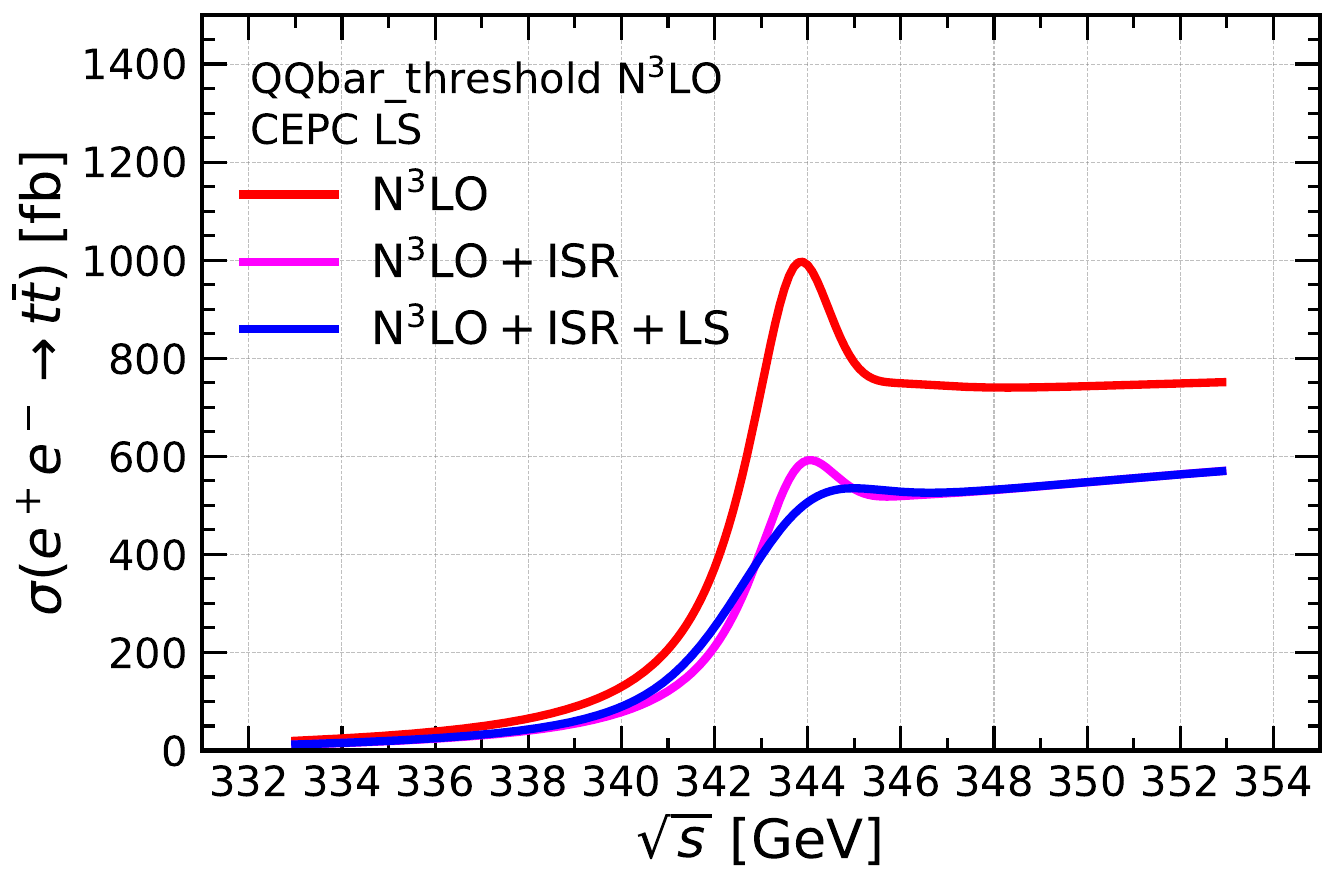}
  \caption{The $\ttbar$ cross-section near threshold as a function of $\sqrts$, 
shown at N$^3$LO, with ISR, and with ISR+LS effects.}
  \label{fig:xs}
\end{figure}

\begin{figure}[htbp]
  \centering
  \includegraphics[scale=0.30]{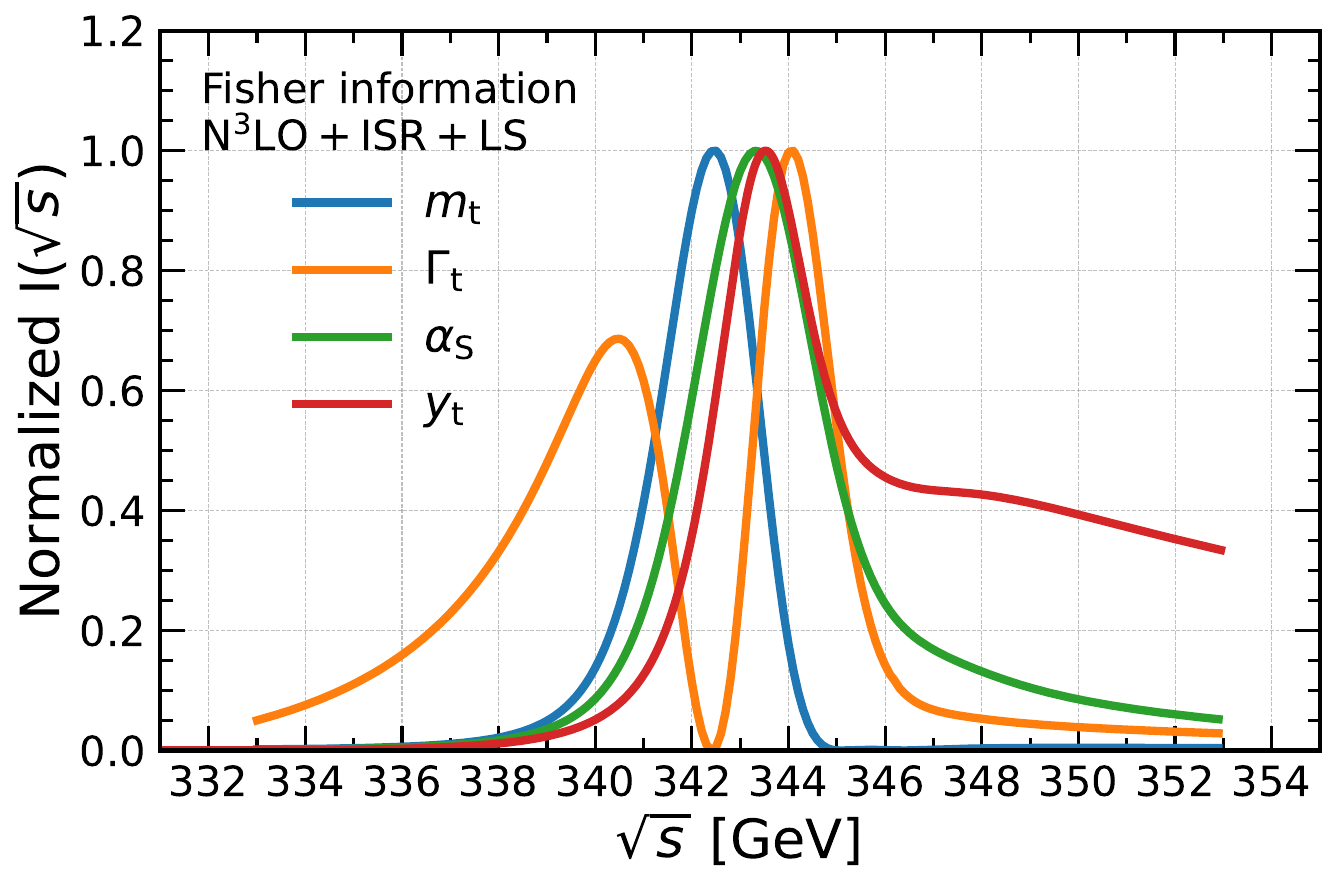}
  \caption{Normalized Fisher information for $\mt$, $\wt$, $\alphas$ and $\yt$ 
as a function of $\sqrts$, based on N$^3$LO+ISR+LS cross-section.}
  \label{fig:fisher}
\end{figure}

As the precision of these measurements is statistically limited, decay channels with large branching ratios are studied: the full-hadronic channel, $e^+e^- \to t\bar{t} \to bq\bar{q}'\bar{b}q''\bar{q}'''$, where both $W$ bosons decay hadronically (BR=44.2\%~\cite{PDG2024}), and the semi-leptonic channel, $e^+e^- \to t\bar{t} \to bq\bar{q}'\bar{b}l\nu$, $l \in \{e, \mu\}$ (BR=29.9\%), where one $W$ decays leptonically (excluding $W\to \tau \nu$).


Monte Carlo samples are generated to evaluate signal efficiencies and background contributions. The \ttbar signal and background samples are generated using MG5\_aMC@NLO 3.6.3~\cite{Alwall:2014hca,Alwall:2014cvc,Artoisenet:2012st,Hirschi:2015iia,Artoisenet:2010cn}, including single top quark, \WW, \ZWW, \ZZ, \ZZZ, \bbbar and \qqbar ($q=u,d,s,c$) backgrounds, with cross-sections listed in Tab.~\ref{tab:xsec_summary}. Events are then showered and hadronized with PYTHIA 8.3~\cite{PYTHIA8.3} and passed through DELPHES~\cite{deFavereau:2013fsa,Selvaggi:2014mya,Mertens:2015kba} for parametric detector simulation and reconstruction.


  \begin{table}[htbp]
    \centering
    \small
    \centering
    \setlength{\tabcolsep}{8pt} 
    \begin{tabular}{llccc}
      \toprule
      Process & Order & \multicolumn{3}{c}{Cross section [fb] at $\sqrt{s}$ [GeV]} \\
      \cmidrule(lr){3-5}
      & & 342.0 & 343.0 & 344.0 \\
      \midrule
      $t\bar{t}$ (full-had) & N$^3$LO & 98.58 & 181.52 & 248.03  \\
      $t\bar{t}$ (semi-lep) & N$^3$LO & 66.75 & 122.89 & 167.93 \\
      Single top            & LO      & 6.01  & 6.40   & 6.80   \\
      $WW$                  & LO      & 11,632 & 11,594  & 11,552  \\
      $ZWW$                 & LO      & 11.05 & 11.27  & 11.47  \\
      $ZZ$                  & LO      & 705.0 & 701.9  & 699.2  \\
      $ZZZ$                 & LO      & 0.608 & 0.605  & 0.603  \\
      $b\bar{b}$            & LO      & 4,678  & 4,646   & 4,621   \\
      $q\bar{q}$            & LO      & 18,486 & 18,381  & 18,258  \\
      \bottomrule
    \end{tabular}
    \caption{Cross-sections of signal and background processes at representative scan energies. The \ttbar signal cross-sections are calculated at N$^3$LO using QQbar\_threshold (v2.2.0)~\cite{Beneke:2016kkb,Beneke:2017rdn}, while backgrounds are computed at LO with MG5\_aMC@NLO 3.6.3.}
    \label{tab:xsec_summary}
  \end{table}

The CEPC DELPHES card~\cite{delphes_cepc} carrying the latest detector description from the reference TDR~\cite{CEPCStudyGroup:2025kmw} is used. Physics objects are reconstructed through a sequence of tracking, calorimetry, particle identification and dedicated jet tagging modules. Electrons, muons and jets, as the basic objects in reconstructing \ttbar events, are introduced in the following.

Electron candidates are reconstructed from tracks with an efficiency parameterized as a function of \pt and $|\eta|$, reaching $\sim$99.7\% for $\pt > 0.5$~GeV within $|\eta| < 3.0$. Tracks are smeared according to the spatial resolution of the vertex detector, silicon tracker and TPC, as well as the time resolution of 30~ps. The electromagnetic calorimeter energy resolution is $\sigma_E / E \approx 1\% \oplus 3\%/\sqrt{E}$ for $|\eta| \leq 3.0$. Candidates with $E > 2$~GeV are reconstructed with 99\% efficiency and must satisfy an isolation criterion: the scalar $\pt$ sum of objects within $\Delta R = 0.5$ must not exceed 12\% of the electron $\pt$.
Muon candidates are reconstructed from tracks using efficiency parameterizations identical to those for electrons. Candidates with $E > 2$~GeV are reconstructed with 99\% efficiency within $|\eta| < 3.0$, identified through inner track matching with the muon spectrometer. Muon isolation is not enforced at the simulation level, as non-prompt muon backgrounds are subdominant.


Jets are reconstructed from particle-flow objects via the $e^+ e^- k_T$ (Durham) algorithm~\cite{Catani:1991hj,Cacciari:2011ma} in an exclusive clustering mode. Events are clustered exclusively into 6 (4) jets for the full-hadronic (semi-leptonic) channel. Jet energies are calibrated using a global correction factor of 1.0075. The $b$-tagging is performed via a parametric algorithm, yielding a 95\% efficiency for $b$-jets and a 1\% mistagging rate for both $c$- and light-flavor jets.


The semi-leptonic channel selects events with the $t\bar{t} \to b\bar{b}q\bar{q}'\ell\nu$ topology: one isolated lepton and 4 jets with $\geq 2$ $b$-tags. To suppress background leptons from jets and $\tau$ decays, a lepton energy threshold of $E > 12$~GeV is applied, as signal leptons from $W$ decays tend to carry higher energy. The impact parameter significance $\text{IPS} = \sqrt{(d_0/\sigma_{d_0})^2 + (d_z/\sigma_{d_z})^2}$, where $d_0$ and $d_z$ are the transverse and longitudinal impact parameters relative to the primary vertex with uncertainties of $\sigma_{d_0}$ and $\sigma_{d_z}$, is required to satisfy IPS $< 3.3$ to reject non-prompt leptons from hadron decays, which exhibit significantly larger IPS values than prompt $W$-decay leptons.


To further suppress backgrounds, a multi-dimensional optimization is performed over several event-level variables by maximizing the statistical significance $Z = \sqrt{2 [(s+b) \ln (1 + s/b) - s]}$, where $s$ and $b$ are the expected signal and background yields. These include the jet resolution parameters ($y_{34}$, $y_{45}$), total and charged particle-flow object (PFO) multiplicities, maximum single-particle momentum, total visible energy, Sphericity and Thrust.

The jet distance parameter in the $e^+e^- k_T$ algorithm~\cite{Catani:1991hj} is defined as $y_{ij} = 2(1-\cos\theta_{ij})\min(E_i^2,E_j^2)/s$, where $E_i$ is the energy of the $i$-th jet, $\theta_{ij}$ the angle between jets $i$ and $j$, and $s$ the center-of-mass energy squared. This dimensionless measure of the clustering cost yields optimal cuts of $\log y_{34} > -3.0$ and $\log y_{45} > -4.3$: signal events with four Born-level partons and hard gluon emissions incur significantly higher costs when clustering from 5 to 4 or 4 to 3 jets than multi-boson and $q\bar{q}$ backgrounds. Signal events also feature higher PFO multiplicities and lower per-particle energy; accordingly, the total and charged PFO multiplicities are required to satisfy $N_{\text{PFO}} \geq 40$ ($N_{\text{PFO}}^{\text{charged}} \geq 14$), the maximum particle momentum $< 86$~GeV, and the total visible energy is required to lie within 212--318~GeV.


Event shape variables Sphericity and Thrust are used to discriminate signal from background. The sphericity tensor $S^{\alpha\beta} = \left(\sum_i p_i^\alpha p_i^\beta / |\mathbf{p}_i|\right) / \left(\sum_i |\mathbf{p}_i|\right)$, where $\alpha,\beta$ denote spatial axes and $p_i^\alpha$ the momentum component of the $i$-th particle, has eigenvalues $\lambda_1 \geq \lambda_2 \geq \lambda_3$, from which the sphericity $S = \frac{3}{2}(\lambda_2+\lambda_3)$ is defined. $S$ ranges from 0 for pencil-like topologies (dijet, diboson) to 1 for sphere-like events ($t\bar{t}$); $S > 0.34$ is required. Thrust, $T = \max_{|\mathbf{n}|=1} [\sum_i |\mathbf{p}_i \cdot \mathbf{n}| / \sum_i |\mathbf{p}_i|]$, where $\mathbf{n}$ is the thrust axis, ranges from 0.5 for isotropic events to 1 for back-to-back dijet topologies; $T < 0.87$ is required.

A kinematic fit is performed to resolve the jet-parton pairing, improve the top-quark mass resolution, and suppress the single-top background that survives the event shape selections. A $\chi^2$ function with 12 constraint terms is constructed, incorporating the conservation of total collision energy ($E_{com}$) and three-momentum components, the $W$ and $t$ mass constraints, and unity scale factors ($sf$) for calibrating the energy of final-state particles:
\begin{equation}
\begin{aligned}
    \chi^2 &= \left( \frac{E_{b_H b_L jj l \nu} - E_{com}}{\sigma_E} \right)^2 + \left( \frac{\Sigma_{b_H b_L jj l \nu}P_{x_i} }{\sigma_{P_x}} \right)^2 \\
    &+ \left( \frac{\Sigma_{b_H b_L jj l \nu}P_{y_i} }{\sigma_{P_y}} \right)^2 + \left( \frac{\Sigma_{b_H b_L jj l \nu}P_{z_i} }{\sigma_{P_z}} \right)^2 \\
    &+ \left( \frac{M_{b_H jj}-M_{t_H}}{\sigma_{M_{t_H}}} \right)^2 + \left( \frac{M_{b_L l \nu}-M_{t_L}}{\sigma_{M_{t_L}}} \right)^2 \\
    &+ \left( \frac{M_{jj}-M_{W_H}}{\sigma_{M_{W_H}}} \right)^2 + \left( \frac{M_{l \nu}-M_{W_L}}{\sigma_{M_{W_L}}} \right)^2 \\
    &+ (sf_{b_H}-1)^2 + (sf_{b_L}-1)^2 + (sf_{jj}-1)^2 + (sf_{l}-1)^2
\end{aligned}
\end{equation}
\noindent
where subscripts $H$/$L$ denote the hadronic/leptonic top decay branches, $b$, $j$, $l$, $\nu$ label the $b$-jet, light-flavor jet, lepton and neutrino. The free parameters in the minimization are the four unity scale factors and the neutrino four-momentum, the latter being uniquely determined by the kinematic constraints. The uncertainties $\sigma$ are taken from simulation. An optimized $\chi^2 < 10$ cut is applied.

The full-hadronic channel ($t\bar{t} \to bq\bar{q}'\bar{b}q''\bar{q}'''$) vetoes leptons using the same criteria as the semi-leptonic channel and requires 6 jets with $\geq 2$ $b$-tags.
The optimized thresholds for the full-hadronic channel are $\log y_{34} > -2.0$, $\log y_{45} > -2.5$, $\log y_{56} > -3.0$, $N_{\text{PFO}} \geq 50$ ($N_{\text{PFO}}^{\text{charged}} \geq 28$), the maximum particle momentum $< 65$~GeV, the total visible energy within 262--359~GeV, Sphericity $> 0.42$ and Thrust $< 0.82$. A kinematic fit is then applied to suppress remaining backgrounds and resolve the pairing of 6 jets. The $\chi^2$ contains 14 constraint terms: energy and four-momentum conservation, top-quark and $W$-boson mass constraints, and six unity scale factors for per-jet energy calibration as the free parameters. An optimized cut of $\chi^2 < 9.0$ is applied.


\begin{align}
\label{eq:lh}
\begin{split}
    \mathcal{L} &= \prod_{i=1}^{N} P( D | (\sigma^{\ttbar}_i(\mt, \wt, \alphas, \yt, \sqrts_i, \xi) \times \epsilon^s_i(\xi) \\
    &+ \sigma^{bkg}_i(\sqrts_i, \xi) \times \epsilon^b_i(\xi))\times L_i(\xi)) \times N(\xi|0,1)
\end{split}
\end{align}
\noindent

The likelihood function $\mathcal{L}$ in Eq.~\ref{eq:lh} is constructed based on the numbers of signal and background events after all selections introduced above, combining a grid of $\sqrts = \{342.0, 342.5, 343.0, 343.5, 344.0\}$~GeV. At each energy point, the observed count $D$ follows a Poisson distribution $P$ whose mean at each scan energy $\sqrts_i$ ($i = 1,\ldots,N$) is the sum of signal and background event yields. The signal cross-section $\sigma^{\ttbar}_i$ depends on the parameters of interest $\mt$, $\wt$, $\alphas$, $\yt$, the colliding energy $\sqrts_i$, and systematic uncertainties $\xi$ constrained by Gaussian priors $N(\xi|0,1)$. The background cross-section $\sigma^{bkg}_i(\sqrts_i, \xi)$, selection efficiencies $\epsilon^{s,b}_i$, and integrated luminosity $L_i$ are evaluated separately at each energy point. The signal efficiencies increase with $\sqrts$, averaged at approximately 60\% and 50\% for the full-hadronic and semi-leptonic channels, respectively, with backgrounds at about 5\% of the total yields.

The impacts of uncertainties at CEPC are analyzed. A multi-dimensional scan is performed on the likelihood as a function of \mt, \wt, \alphas and \yt, while all systematic uncertainties except the theoretical one on the \ttbar cross-section are profiled. The total and breakdown uncertainties in \mt and \wt measurements are presented in Tab.~\ref{tab:uncert}.

The leading systematic uncertainty arises from the top Yukawa coupling $\yt$, expected to reach 3\% by the HL-LHC era~\cite{Azzi:2019yne} and to be further constrained above the $\ttbar$ threshold, inducing cross-section variations of about 0.5\% and impacts of 4.2~MeV and 5.9~MeV on $\mt$ and $\wt$, respectively. The second largest contribution is from $\alphas$, projected to reach 0.0001 or better using $Z$-boson observables~\cite{dEnterria:2020cpv} and energy correlations~\cite{Lin:2024lsj}, varying the cross-section by about 0.25\% with impacts of 2.4~MeV and 3.1~MeV on $\mt$ and $\wt$, respectively. The luminosity spectrum uncertainty, conservatively assumed at 1\% consistent with FCC-ee~\cite{Fcc-ee2025}, affects the cross-section by less than 0.1\% and is the third leading source with 2.8~MeV impact on $\wt$, as it smears the cross-section enhancement peak whose width is highly sensitive to $\wt$. The beam energy uncertainty of 1.8~MeV per beam~\cite{Tang:2020gmv} affects the cross-section by 1\% or less, impacting $\mt$ by 1.3~MeV with negligible effect on $\wt$. The $b$-tagging uncertainty of 1\%~\cite{Fcc-ee2025} contributes 1.0~MeV and 1.7~MeV to $\mt$ and $\wt$, while the luminosity measurement uncertainty of 0.01\% from the $Z$ pole~\cite{CEPCStudyGroup:2025kmw} has a trivial impact of 0.1~MeV.

The theoretical uncertainty of \ttbar cross-section is evaluated separately without profiling, by varying the renormalization scale by factors of 0.5 and 2.0~\cite{Beneke:2024sfa}. The envelope, taken as the uncertainty by convention, results in highly asymmetric cross-section variations with a marginal upward error but a significant downward shift of up to $-5\%$. The corresponding impacts on $\mt$ and $\wt$ are $^{+1.3}_{-32.9}$~MeV and $^{<+0.1}_{-79.9}$~MeV, respectively.

In summary, excluding the theoretical cross-section uncertainty, the total uncertainties on $\mt$ and $\wt$ are 7.5~(6.8) and 19.4~(16.2)~MeV in the BASE (EXT) scenario, with respect to the central values of $\mt=171.5$~GeV and $\wt=1.33$~GeV, dominated by the statistical uncertainties of 5.6~(4.6) and 18.0~(14.6)~MeV. The theoretical uncertainty can further impact $\mt$ by 32.9~MeV and $\wt$ by 79.9~MeV, highlighting significant room for improvement with higher-order calculations. The two-dimensional likelihood scans of $\mt$ and $\wt$ are presented in Fig.~\ref{fig:2d-56} and Fig.~\ref{fig:2d-84} for the BASE and EXT scenarios, respectively.

Additionally, likelihood scans are performed to extract $\alphas$ and $\yt$ at the threshold, with $\mt$ and $\wt$ profiled as free parameters. Excluding the theoretical uncertainty of \ttbar cross-section, the total uncertainties are $\delta\alphas = 0.00143~(0.00116)$ and $\delta\yt = 0.197~(0.160)$ in the BASE (EXT) scenario, dominated by the statistical uncertainties of 0.00141~(0.00113) and 0.196~(0.159). The theoretical uncertainty induces a large additional uncertainty of 0.00343 on $\alphas$ and 0.453 on $\yt$.

In conclusion, using the latest CEPC reference detector design for the first time, we demonstrate that the $\ttbar$ threshold-scan method can simultaneously measure $\mt$, $\wt$, $\alphas$ and $\yt$. The precision of all measurements is currently dominated by the theoretical uncertainty of \ttbar cross-section, computed up to N$^3$LO. If improved to the level of the statistical uncertainty, the top-quark mass precision can reach a few MeV --- nearly two orders of magnitude beyond the HL-LHC, corresponding to a relative precision of $10^{-5}$ --- and the top-quark width precision improves by one order of magnitude. The $\alphas$ and $\yt$ measurements are complementary to those from the $Z$ pole and above the $\ttbar$ threshold, respectively. Such measurements at the CEPC --- or, equivalently, the FCC-ee --- would enable a decisive advance in precision tests of the Standard Model and provide a definitive probe of the electroweak vacuum stability.

\begin{table}[!ht]
    \centering
    \begin{tabular}{lcccc}
    \hline
        Uncertainties & \multicolumn{2}{c}{\mt [MeV]} & \multicolumn{2}{c}{\wt [MeV]} \\
        ~ & BASE & EXT & BASE & EXT  \\
        \hline
        Statistics & $\pm 5.6$  & $\pm 4.6$  & $\pm 18.0$  & $\pm 14.6$  \\
        \hline
        \alphas & \multicolumn{2}{c}{$\pm 2.4$} & \multicolumn{2}{c}{$\pm 3.1$}   \\ 
        \yt & \multicolumn{2}{c}{$\pm 4.2$} & \multicolumn{2}{c}{$\pm 5.9$}  \\ 
        Luminosity & \multicolumn{2}{c}{$\pm 0.1$} & \multicolumn{2}{c}{$\pm 0.1$} \\ 
        BE & \multicolumn{2}{c}{$\pm 1.3$} & \multicolumn{2}{c}{$<\pm 0.1$} \\ 
        LS & \multicolumn{2}{c}{$\pm 0.2$} & \multicolumn{2}{c}{$\pm 2.8$}  \\ 
        Btagging & \multicolumn{2}{c}{$\pm 1.0$} & \multicolumn{2}{c}{$\pm 1.7$}  \\
        \hline
        Total & $\pm 7.5$  & $\pm 6.8$  & $\pm 19.4$  & $\pm 16.2$   \\
        Theory & \multicolumn{2}{c}{$^{+1.3}_{-32.9}$} & \multicolumn{2}{c}{$^{<+0.1}_{-79.9}$} \\ \hline
    \end{tabular}
    \caption{Uncertainties. BE stands for beam energy, while LS for luminosity spectrum.}
    \label{tab:uncert}
\end{table}

\begin{figure}[htbp]
  \centering
  \includegraphics[scale=0.27]{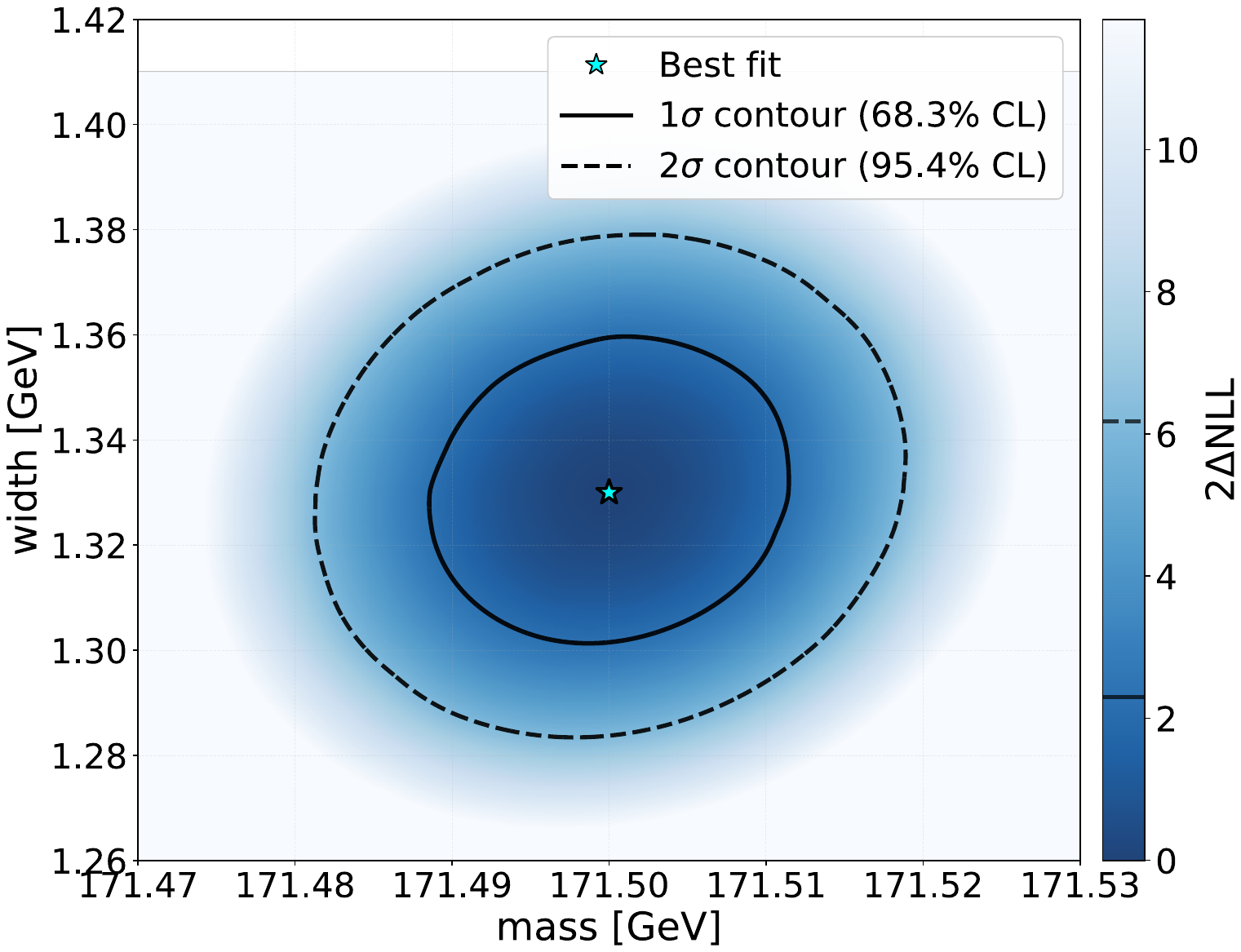}
  \caption{Two-dimensional likelihood scan over \mt and \wt with full systematic uncertainties except the theoretical one from the \ttbar cross-section calculation, in the BASE scenario using 5 energy points with 56 \ifb per point.}
  \label{fig:2d-56}
\end{figure}

\begin{figure}[htbp]
  \centering
  \includegraphics[scale=0.27]{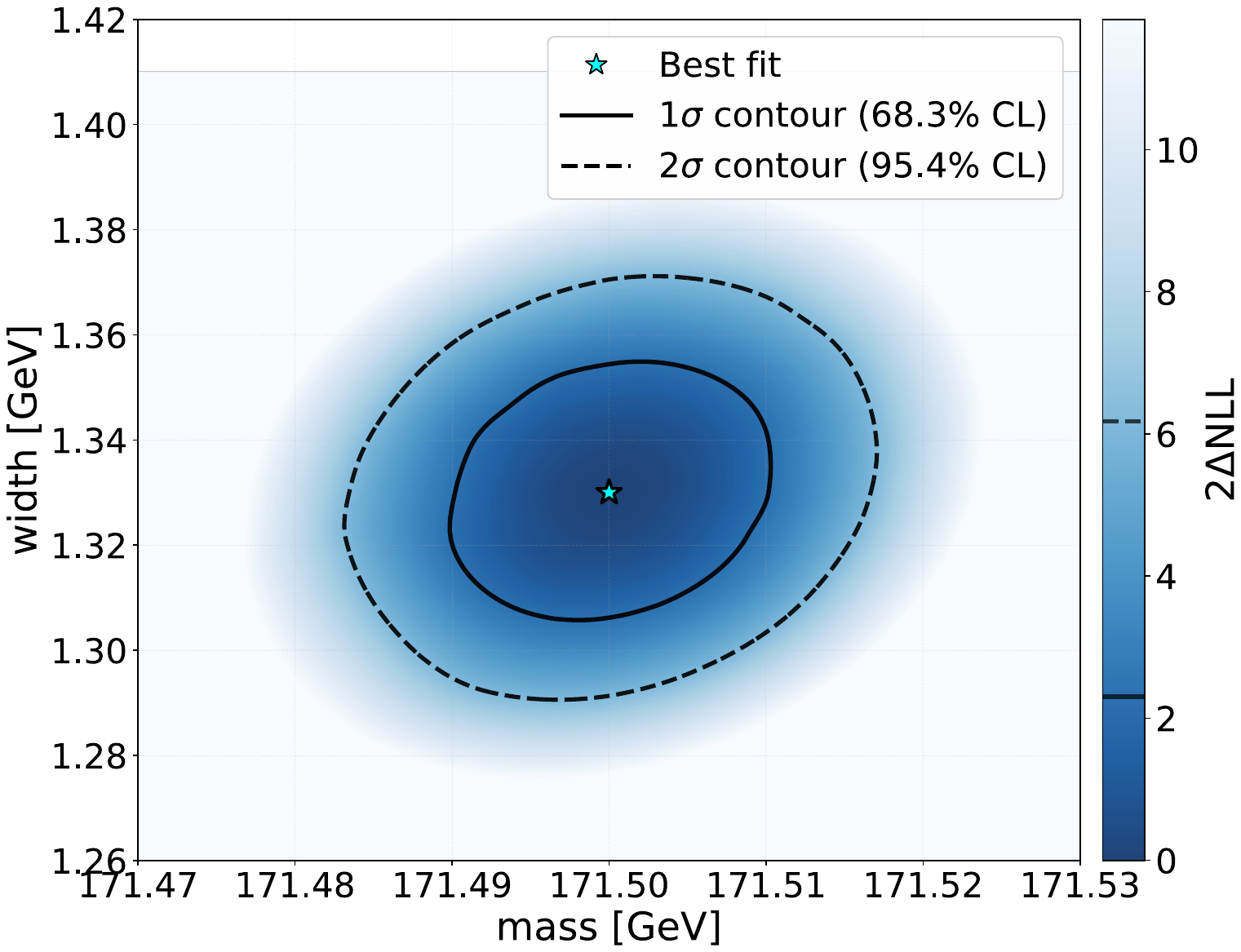}
  \caption{Two-dimensional likelihood scan over \mt and \wt with full systematic uncertainties except the theoretical one from the \ttbar cross-section calculation, in the EXT scenario using 5 energy points with 84 \ifb per point.}
  \label{fig:2d-84}
\end{figure}

\FloatBarrier

\begin{acknowledgments}
We thank our CEPC colleagues for helpful discussions and technical support. This work is supported in part by the National Science Foundation of China under Grants No. 12188102.
\end{acknowledgments}


\bibliography{main_new_52}

\end{document}
%